# Polymer Genome in the Age of Artificial Intelligence

*Jifeng Wang*[1], *Ying Wang*[1]*

1. State Key Laboratory of Molecular Engineering of Polymers
   Department of Macromolecular Science
   Research Center of AI for Polymer Science
   Fudan University
   Shanghai, 200438, China

*Corresponding author email: wying@fudan.edu.cn

**ABSTRACT:** Artificial intelligence (AI) is redefining the landscape of polymer science. Although numerous AI applications have been introduced in this field, the roles of polymer encoding strategies and different applications of AI models in polymer design remain insufficiently understood. Here, we build upon the foundation of polymer databases to critically examine the performance and applicability of current encoding strategies across different use cases. We then focus on two major AI application domains, property prediction and inverse design, to evaluate the strengths, weaknesses and suitable scenarios for various model architectures. Finally, we emphasize the significance of the online platforms for promoting data accessibility and accelerating the migration from experience-based discovery toward AI-driven innovation in polymer science and engineering. Through these discussions, we aim to provide practical guidance for future research and development in AI-assisted polymer design.

**Keywords:** Polymer Genome, Polymer Database, Polymer Encoding Strategy, Polymer Prediction Model, Polymer Generative Model

## INTRODUCTION

Artificial intelligence (AI) is transforming the landscape of polymer science by accelerating the shift from empirical exploration toward intelligent and data-driven discovery.[1, 2] However, compared with the rapid development of AI in molecules and crystals, the development of AI-driven research on polymers has been left behind. This lag primarily stems from the inherent structural complexity of polymers, including their large chemical design space and multi-scale architectures.[3, 4]

In recent years, researchers have introduced a range of powerful tools to address these challenges.[5, 6] The gradual emergence of open, standardized and scalable polymer databases has laid a foundation for high-quality model training. Meanwhile, polymer encoding strategies have evolved from hand-crafted descriptors and SMILES[7] strings to more expressive encodings such as 2D/3D molecular graphs and sequence representations. Correspondingly, AI model architectures are undergoing a paradigm shift. Traditional machine learning models based on hand-crafted descriptors are being replaced by deep learning models, including graph neural networks (GNNs)[8, 9], Transformer-based architectures[10-12] (e.g., BERT[13], GPT[14]), and generative models such as variational autoencoders (VAE)[15] and denoising diffusion probabilistic models (DDPM)[16]. These advanced models not only facilitate the inverse design of polymers but also enable a deeper understanding of structure-property relationships. Moreover, the rise of integrated online platforms has substantially improved the accessibility, usability, and reproducibility of AI tools in polymer science. Through user-friendly interfaces, these platforms allow researchers to conveniently apply AI tools in polymer research.

Despite recent progress, the field of AI-driven polymer science still faces critical challenges, particularly in the availability of high-quality data[17, 18] and the development of unified encoding adaptable to both homopolymers and copolymers[19, 20]. In addition, although numerous encoding strategies and models for polymers have been proposed, comprehensive comparisons of their applicable scenarios, advantages and limitations remain scarce. In this review, we aim to address this gap by discussing the limitations of polymer databases, encoding strategies, models, and online platforms, thereby providing practical guidance for future research.

## POLYMER DATABASE

Data are the foundation of AI-driven polymer discovery. In the context of polymer science, several dedicated databases have been developed in recent years. As shown in **Figure 1 (a)**, the evolution of polymer databases reveals a clear trend toward openness, standardization and scalability.

Early efforts such as PolyInfo[21] laid the foundation for machine learning of polymer science. Integrating over $5.5 \times 10^5$ polymer-property pairs for homopolymers, copolymers and polymer blends (**Figure 1 (b)**), PolyInfo provides extensive information such as polymer thermodynamic and mechanical properties, making it the earliest large-scale dataset for model training in the field of polymer science. However, its accessibility restriction highlights the long-standing issue of data openness in polymer research. Subsequent database such as Khazana[22] complemented the experimental data with electronic and dielectric properties obtained via first-principles calculations, though the accuracy of the calculated dataset remains to be validated.

To overcome the limitations of data scarcity, several virtually generated databases, including PI1M[23], Open Macromolecular Genome (OMG)[24] and SMiPoly[25], have been established. These databases expand the structural diversity through algorithmic or reaction-template-based polymer generation while embedding synthetic feasibility constraints. For example, **Figure 1 (c)** compares data distributions between the OMG and PolyInfo, showing that virtually generated polymers can significantly expand the structural space while maintaining the realistic chemical distributions. However, virtually generated databases focus mainly on structural diversity rather than property annotations, making them suitable only for unsupervised learning or large-scale screening.

The emergence of large language models (LLMs) has recently transformed the acquisition of polymer datasets. Gupta et al.[26] employed models such as MaterialsBERT[27] and GPT-3.5 to automatically extract polymer-property pairs from scientific literature, aggregating about $3.9 \times 10^5$ records and demonstrating the feasibility of large-scale text-mined datasets. However, automated extraction introduces redundancy and inconsistency, for example, multiple entries referring to the same polymer may report different measured values or experimental conditions. To address this challenge, OpenPoly[28] integrates LLM-based extraction with rule-based curation and employs a 50% trimmed-mean, ultimately producing $2.1 \times 10^4$ high-confidence entries (**Figure 1 (d)**). The curated data are publicly hosted on the CEMP[29].

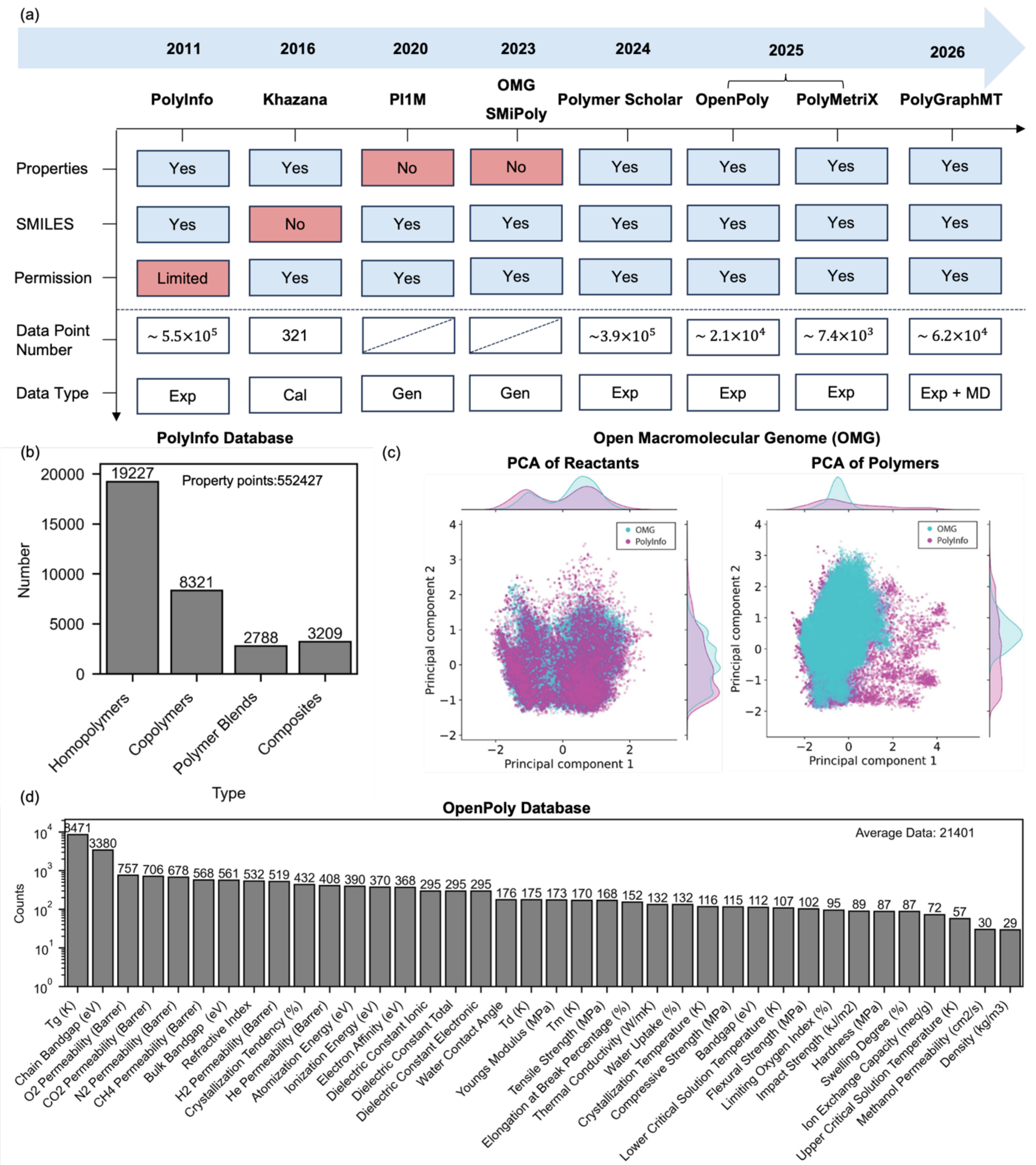

**Figure 1**. **Overview of representative polymer databases**[21-26, 28, 30, 31]**.** (a) Comparison of data coverage and features among main polymer databases in the field. “Properties” indicates whether the database includes experimental or computational property data; “SMILES” denotes the availability of polymer structures encoded as SMILES strings; “Permission” reflects the degree of data accessibility; and “Data Type” distinguishes experimentally measured (“Exp”), theoretically calculated (“Cal”), and virtually generated (“Gen”) data. (b) Distribution of polymer types and property data points in the PolyInfo database. (c) Principal component analysis (PCA) projection of reactants and polymers from the OMG and PolyInfo databases. (d) Data points across different property categories within OpenPoly database. Panel (c) reproduced from ref.[24] under the CC BY-NC-ND 4.0 license.

Beyond simply increasing dataset size, recent work in polymer informatics has placed greater emphasis on chemically meaningful structural representations and the integration of heterogeneous data sources. For example, PolyMetriX[30] couples a curated dataset of 7,367 unique $T_g$ data points with a hierarchical featurization scheme that separately describes the full polymer representation, backbone, and side chains, thereby providing a more interpretable connection between molecular architecture and glass-transition behavior. In parallel, ADEPT–PolyGraphMT[31] integrates experimental measurements with data from molecular dynamics (MD) simulations, density functional theory (DFT) calculations, and group-contribution estimates, yielding a unified multi-fidelity dataset of approximately 62,000 values across 28 properties. The study further shows that lower-fidelity computational data, despite source-dependent systematic biases, can retain meaningful global structure–property trends and broaden chemical-space coverage.

Despite the ongoing efforts to refine the polymer databases, we still face the challenges of limited records, uneven quality and incomplete representation of the multiscale structures of polymers. Experimental datasets remain fragmented and lack standardized protocols for testing and representation. For calculated data, the periodic structure and complexity of polymer topologies only allow limited electronic properties[32] based on the DFT calculations; whereas, the key properties such as $T_g$ and mechanical performance still rely on the computationally expensive MD simulations whose accuracy only depends on potential functions and sampling adequacy.[33, 34]

Future polymer databases should therefore move beyond simple polymer–property pairs. Priority should be given to linking each reported property value to measurement protocol and experimental conditions and recording polymer-specific identity descriptors, (such as the degree of polymerization and molecular-weight distribution). Historically, curating such information has required substantial manual effort and domain expertise, creating a considerable entry barrier for newly established research teams. Recent advances in large language models (LLMs) and agentic workflows are beginning to lower this barrier. Domain-specific language models and proprietary GPT-based systems have already enabled the large-scale extraction, normalization, and organization of polymer–property records from scientific literature while substantially reducing the need for manually defined extraction rules and extensive task-specific annotation.[35, 36] These developments indicate that LLM-based agents can support the efficient construction and continuous updating of structured polymer databases, provided that their outputs are constrained by predefined schemas and expert validation.

## POLYMER ENCODING STRATEGY

Polymer encoding strategy plays a key role in enabling machine learning models to interpret and learn from polymer structures. The choice of encoding directly influences model's ability to extract meaningful chemical patterns, impacting both predictive accuracy and design interpretability. Reliable polymer design depends on high-quality data, as well as efficient encodings of the polymer structure. Over the past five years, a variety of polymer encoding strategies have been proposed. These methodologies can be generally classified into fingerprint-based, graph-based, or sequence-based encodings as listed in **Table 1**. Each strategy captures chemical information at a different level of detail.

**Table 1.** Common polymer encoding strategies and representative models for property prediction and generative tasks

| Representation | Encoding Strategy | Task Type | Hand-crafted Descriptors | Representative Models |
|---|---|---|---|---|
| Morgan fingerprint[37] | Fingerprint | Regression | Not required | XGBoost, MLP |
| Polymer Genome fingerprint[38] | Fingerprint | Regression | Required | XGBoost, MLP |
| PolyBERT fingerprint[39] | Fingerprint | Regression | Not required | XGBoost, MLP |
| 2D molecular graph | Graph | Regression/ Generation | Node and edge features | 2D GNN (GAT, GCN) |
| 3D molecular graph | Graph | Regression/ Generation | Node and edge features | 3D GNN (EGNN, IGNN) |
| SMILES one-hot encoding | Sequence | Generation | One-hot vocabulary | GAN, VAE, DDPM |
| SMILES sequence | Sequence | Regression/ Generation | Tokenization rules | Transformer |

**Fingerprint-based encoding strategies**. Fingerprints are widely used for property prediction due to their computational efficiency. Morgan fingerprints, which can be generated directly through RDKit[40], provide a fast and convenient approach for regression tasks across multiple properties. Polymer Genome fingerprints[38] are hand-crafted descriptors that encode multiscale polymer features, including atomic, bonding and chain-level information, providing chemically interpretable representations for machine learning tasks. PolyBERT fingerprints[39], trained through self-supervised learning on large polymer SMILES datasets, captures latent polymer semantics and performs robustly in transfer-learning scenarios. A recent study has shown that under data-scarce conditions, using Fingerprint-based encoding strategies can outperform deep learning architectures, such as GNN.[28]

Despite their efficiency, these strategies share a common limitation: they struggle to encode global structural features such as polymerization degree, copolymer sequence, or compositional distribution.

**Graph-based encoding strategies**. Graph-based encoding represents polymers as networks of nodes and edges, providing a flexible framework for capturing connectivity and chemical bonding information, making it suitable for both property prediction and generative modeling tasks. 2D molecular graphs encode atomic connectivity and bond types, suitable for graph convolutional networks (GCNs) and graph attention networks (GATs). 3D molecular graphs incorporate geometric parameters such as bond lengths, bond angles and dihedral angles, often used with equivariant (EGNN) or invariant (IGNN) architectures that respect molecular symmetry. Graph methods excel at learning local and global structural interactions and have advanced understanding of structure-property relationships. However, they often require task-specific feature engineering for node and edge attributes and typically demand larger training datasets.

**Sequence-based encoding strategies**. Sequence representations offer strong potential for sequential modeling of polymers. Canonical formats such as SMILES and BigSMILES[41] are commonly used to encode polymer structures as character sequences. These sequences can be augmented with additional information, such as polymerization degree or processing conditions, making sequence-based encodings highly extensible and adaptable to diverse modeling tasks. Owing to their compatibility with natural language processing techniques and their ability to incorporate polymer information directly into the input, sequence-based encodings have been successfully applied to both property prediction and polymer generation tasks. However, this strategy typically requires larger training datasets to enable models to effectively learn the underlying syntax and structural rules embedded in sequence formats such as SMILES.

In summary, fingerprint-based encodings are fast and convenient for property prediction; graph-based encodings capture structural topology and local interactions in detail; and sequence-based approaches provide flexibility and semantic depth, especially for complex and multicomponent systems. **Figure 2** illustrates the common polymer representations for homopolymers (**Figure 2 (a)**) and copolymers (**Figure 2 (b-d)**). As each encoding strategy presents distinct advantages and limitations, current research is shifting toward integrating multiple encodings to improve model robustness and generalization. For instance, Huang et al.[42] proposed the Uni-Poly framework, which integrates fingerprints, 2D/3D graphs, and SMILES sequences, outperforming single-modality models in predicting multiple properties such as $T_g$ and $T_m$. Although encoding strategies for homopolymers have been well established, the encoding of copolymers remains an open challenge due to compositional complexity, variable sequences and uncertain repeating units. Recent work by Kuenneth et al.[43] linearly accumulated the Polymer Genome fingerprints by monomer fraction to represent copolymers (**Figure 2 (b)**). However, it cannot efficiently distinguish copolymers with different sequences. Jiang et al.[44] introduced a 2D graph-based encoding method (**Figure 2 (c)**) that represents entire polymer chains. This approach accounts for sequence information but requires prior knowledge of how the repeating units are connected. In contrast, Xu et al.[45] employed a Transformer model using SMILES sequences and compositional ratios (**Figure 2 (d)**) to achieve unified copolymer encoding. Although this method incorporates richer compositional information, it still fails to capture the intrinsic sequence-dependence in copolymers.

Automated inference of copolymer sequences has recently been demonstrated for selected polymerization mechanisms. Zhang et al.[46] introduced reactivity-ratio fingerprints (rFPs) and used deep learning to infer reactivity ratios from sparse experimental data across binary and ternary copolymerization and under varied solvent and temperature conditions, then the inferred kinetics were used to guide sequence-distribution control. Vogel et al.[47] developed a semi-supervised variational autoencoder for copolymer representation. A weighted directed message-passing network encoded atomistic monomer graphs, with monomer fractions represented by node weights and intermonomer connection probabilities represented by stochastic edge weights. A Transformer decoder then generated the corresponding monomer SMILES, composition and connectivity rules, enabling the representation of random, block, and alternating copolymers. Thus, the remaining frontier is a transferable representation that connects polymerization mechanism and conditions to sequence distributions, molecular-weight distributions and complex topology while remaining compatible with property prediction and generative design.

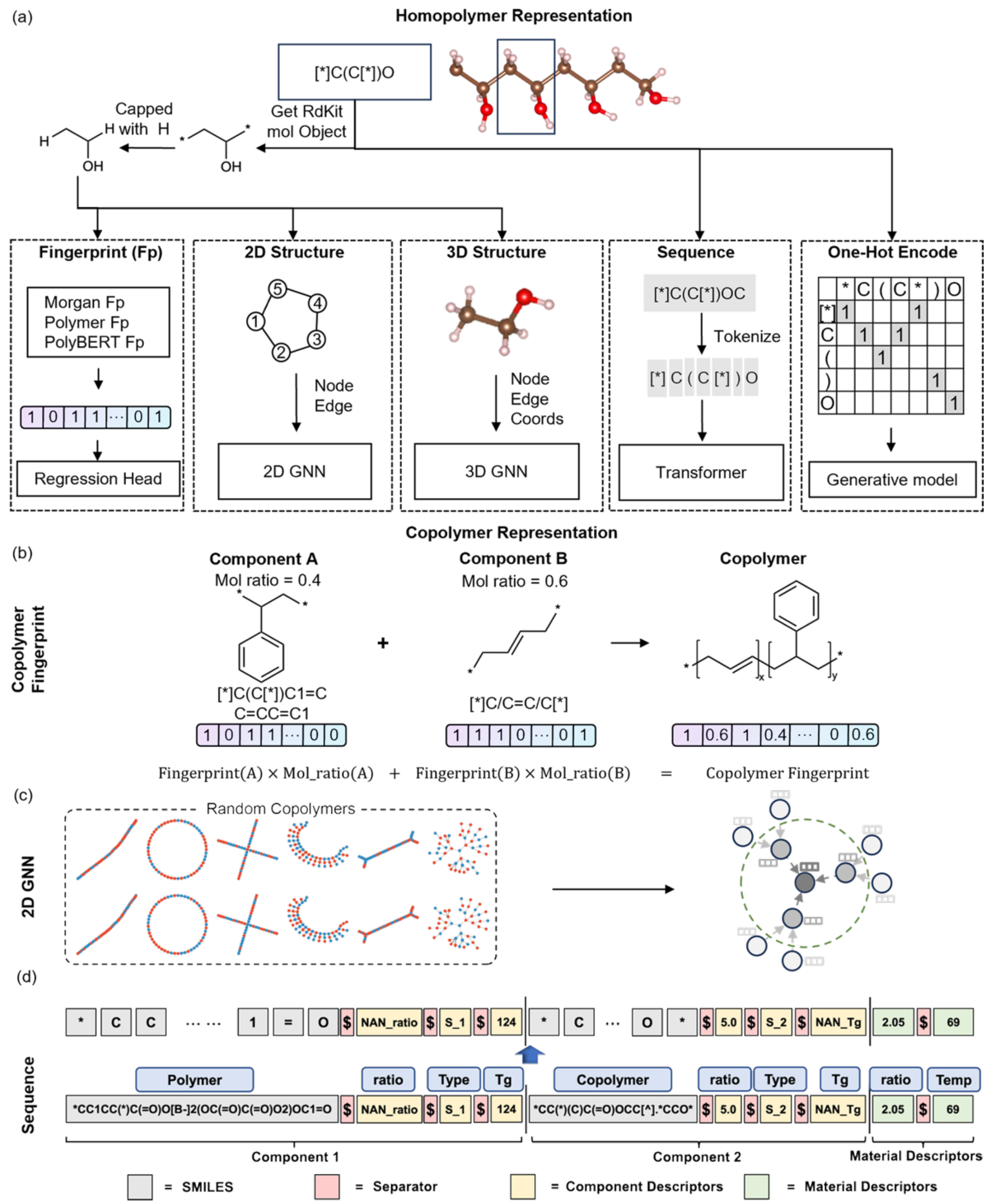

**Figure 2**. **Polymer encoding strategies for property prediction and generation**. (a) Encoding of homopolymers. Common structural representations used in predictive and generative modeling, including repeat-unit–based fingerprints, 2D/3D molecular graphs, SMILES sequence, and SMILES one-hot encodings. (b–d) Encoding of copolymers. (b) Copolymer fingerprints generated by combining monomer fingerprints according to their molar ratios. (c) When the copolymer sequence is known, 2D molecular graph encoding can be employed to capture interunit connectivity. (d) When the sequence is unknown, a sequence-based representation can be used to embed both the monomer components and their molar ratios for flexible structural representation and property prediction of arbitrary copolymers. Panel (c) adapted with permission from ref. [44]. Copyright 2025 American Chemical Society. Panel (d) reproduced from ref. [45] under the CC BY-NC-ND 4.0 license.

Finally, the choice of encoding strategy should align with the data availability, prediction target and structural complexity of the polymer system. When rapid property prediction is desired under data-scarce conditions, fingerprint-based encoding strategies are preferred. For tasks requiring automatic extraction of structure-property relationships, or when fingerprints are insufficient or unavailable, particularly for node-level predictions, graph-based encoding strategies are more appropriate. Finally, sequence-based encoding strategies provide a flexible alternative for integrating diverse external information.

## POLYMER MODELING

### PROPERTY PREDICTION MODELS

Property prediction is the primary application of AI in the field of polymer science.[48] This section outlines the development of AI models for polymer property prediction and discusses their current limitations and future research directions.

In the early stage of AI-assisted polymer property prediction, research relied on hand-crafted molecular descriptors and fingerprints, as illustrated in **Figure 3 (a)**. A notable example is the Polymer Genome platform[38], developed by Ramprasad in 2018. This platform constructed polymer features using hand-crafted descriptors and employed kernel ridge regression to predict properties such as $T_g$, achieving accuracies comparable to experimental uncertainty. While such descriptor-based machine learning models offer strong interpretability, they lack the ability to automatically extract features from input data. With the rise of deep learning, the field entered a new phase focused on automatic feature extraction from molecular structure. Researchers began using GNN and Transformer-based architectures to directly learn representations from polymer structure. For instance, by encoding polymer repeating unit as node and edge features (**Figure 3 (b)**), GNN has been used to predict thermal, mechanical and electronic properties for polymers.[49-52] These studies demonstrated that GNN can effectively capture the complex relationships between polymer structure and performance. However, there remain challenges including the varied lengths and sequences in polymer chains, making it difficult to construct complete molecular graphs. Representations based solely on repeating unit omit essential information such as polymerization degree and interchain interactions.[53]

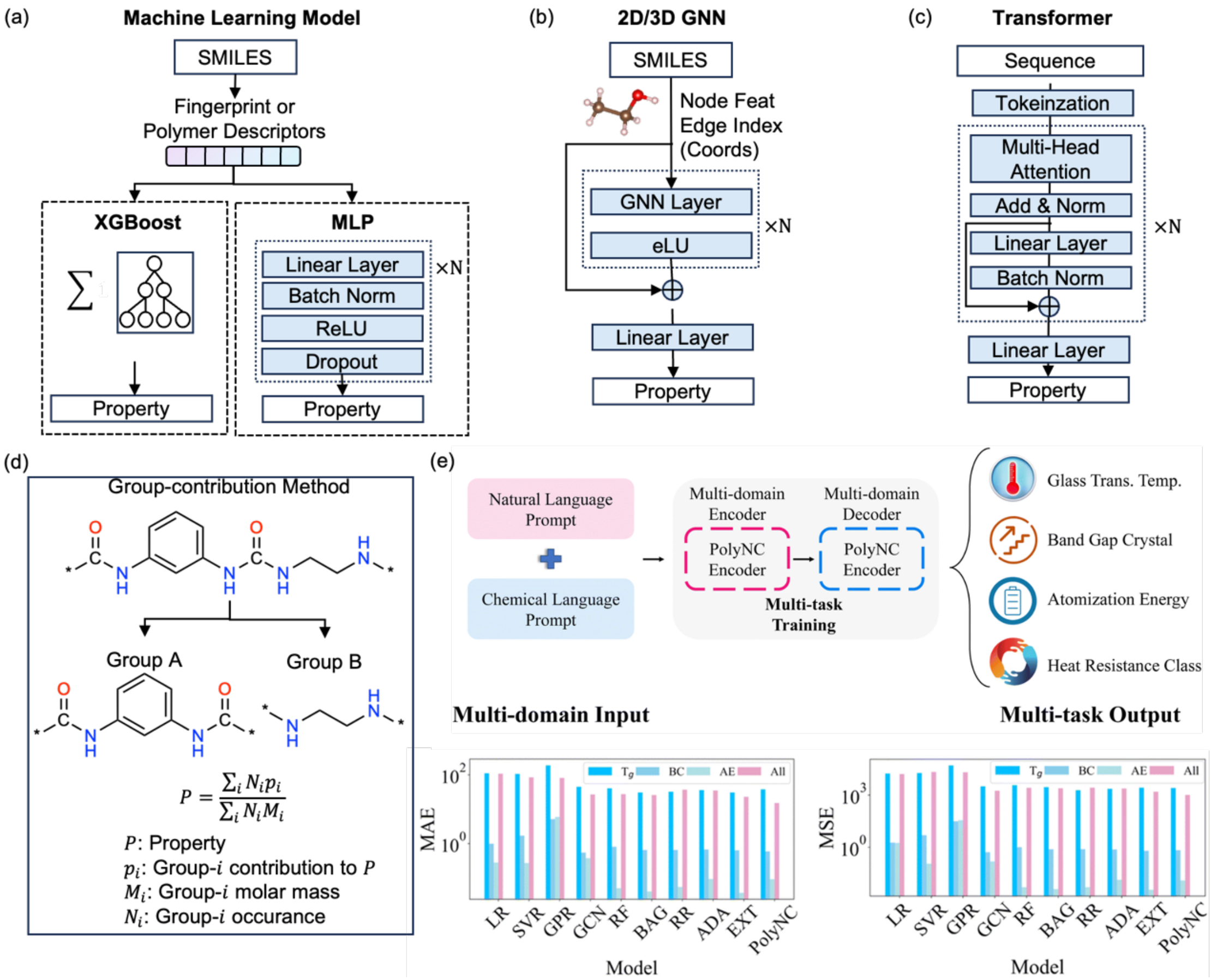

**Figure 3**. **Overview of AI models for polymer property prediction.** (a) Traditional machine learning workflow. Polymer structures are represented by SMILES-derived fingerprints or hand-crafted descriptors, followed by property prediction using models such as XGBoost or MLP. (b) Graph neural network (GNN) models. Polymers are encoded as graphs defined by node features and edge indices representing monomer connectivity. (c) Transformer-based models. Sequence inputs (e.g., SMILES of repeating unit) are tokenized and processed through a multi-head attention mechanism to learn inter-sequence correlations for property prediction. (d) Data augmentation under limited datasets. Group contribution methods are used to estimate polymer properties and expand labeled datasets for model training. (e) PolyNC model. A Transformer-based architecture for polymer property prediction. Panel (e) reproduced from ref.[54] under the CC BY-NC 3.0 license.

The introduction of the Transformer architecture has opened new opportunities for polymer property prediction (**Figure 3 (c)**). In contrast to GNN, which primarily model local structural relationships, Transformer offers greater input flexibility, enabling the incorporation of additional contextual information such as polymerization degree. Xu et al. [45] proposed TransPolymer, a representative example in this category. The model encodes polymer SMILES sequences along with compositional descriptors, such as monomer ratios, to capture multiscale information. It employs masked language modeling pretraining on large unlabeled datasets followed by multitask fine-tuning, enabling robust property prediction across homopolymers and copolymers within a unified framework. This work marked a transition from structural recognition to semantic understanding of polymers and laid the foundation for transferable and interpretable polymer foundation models.

Transformer-based models typically require large volumes of data to train effectively,[55] which poses a significant challenge in polymer science due to the limited availability of high-quality datasets. To address this limitation, Liu et al.[56] constructed high-confidence polymer datasets using a group-contribution strategy (**Figure 3 (d)**), expanding the accessible chemical space and improving model generalization. Then, Qiu et al.[54] introduced PolyNC, a Text-to-Text (T5)[57] framework that combines natural language prompts with polymer SMILES as inputs to jointly predict properties such as $T_g$ and electronic characteristics (**Figure 3 (e)**). This chemical–language hybrid approach enhances human-AI interaction, offering a new paradigm for polymer design. Overall, two complementary directions have emerged for polymer property prediction. GNN-based models emphasize physically grounded structural representations, while Transformer-based models focus on global semantic learning and the ability to encode complex structural and contextual information. GNNs remain closer to chemical realism, but are limited by the structural complexity of polymers and often fail to capture the full extent of polymer chains. In contrast, Transformers achieve broader generalization but require large datasets and struggle to account for repeating unit sequence.

Beyond predictive accuracy, the reliability of model predictions is receiving increasing attention. Conventional metrics such as mean absolute error (MAE) and coefficient of determination ($R^2$) describe performance on test set, but they provide little information about the model's applicability domain or its reliability in unexplored chemical space. Tang et al.[58] benchmarked nine uncertainty-quantification methods across four polymer properties, including $T_g$, band gap, melting temperature ($T_m$), and thermal decomposition temperature ($T_d$). The evaluation included out-of-distribution experimental datasets and polymers from different chemical classes. The results showed that no single method performed consistently best across all settings. Ensemble methods were generally robust for in-distribution predictions, whereas Bayesian or boosting-based approaches were more effective in certain out-of-distribution cases. These findings show that conventional accuracy metrics alone are insufficient. Reliable model evaluation should also consider uncertainty calibration, the agreement between predicted uncertainty and actual error, and the boundaries of the applicability domain. The origin and fidelity of the underlying data must also be taken into account.

## GENERATIVE MODELS

The development of generative models has transformed polymer research from property prediction to inverse design. These models learn the distribution of chemical space to generate polymer structures that meet target properties, offering new pathways for high-throughput screening and materials discovery. Early work explored generative adversarial networks (GANs) [59], an important branch of generative learning, but their application to polymer SMILES generation proved challenging due to unstable training and difficulty ensuring chemical validity.[60] As a result, VAE, DDPMs and Transformer-based architectures have become the dominant approaches for polymer generation. VAE learned a continuous latent space that enabled smooth interpolation between structures and property-guided design, establishing a framework for target-oriented molecular generation. Building upon VAE, DDPM improved the stability and diversity of generated structures by learning to iteratively reconstruct molecules from noise, thereby allowing broader exploration of chemical space for polymers. More recently, Transformer-based models have introduced a language-based approach to polymer generation. Leveraging autoregressive learning, they can directly generate polymer structures conditioned on target properties, thereby enabling semantics-driven molecular design. **Figure 4** illustrates representative architectures of these generative models.

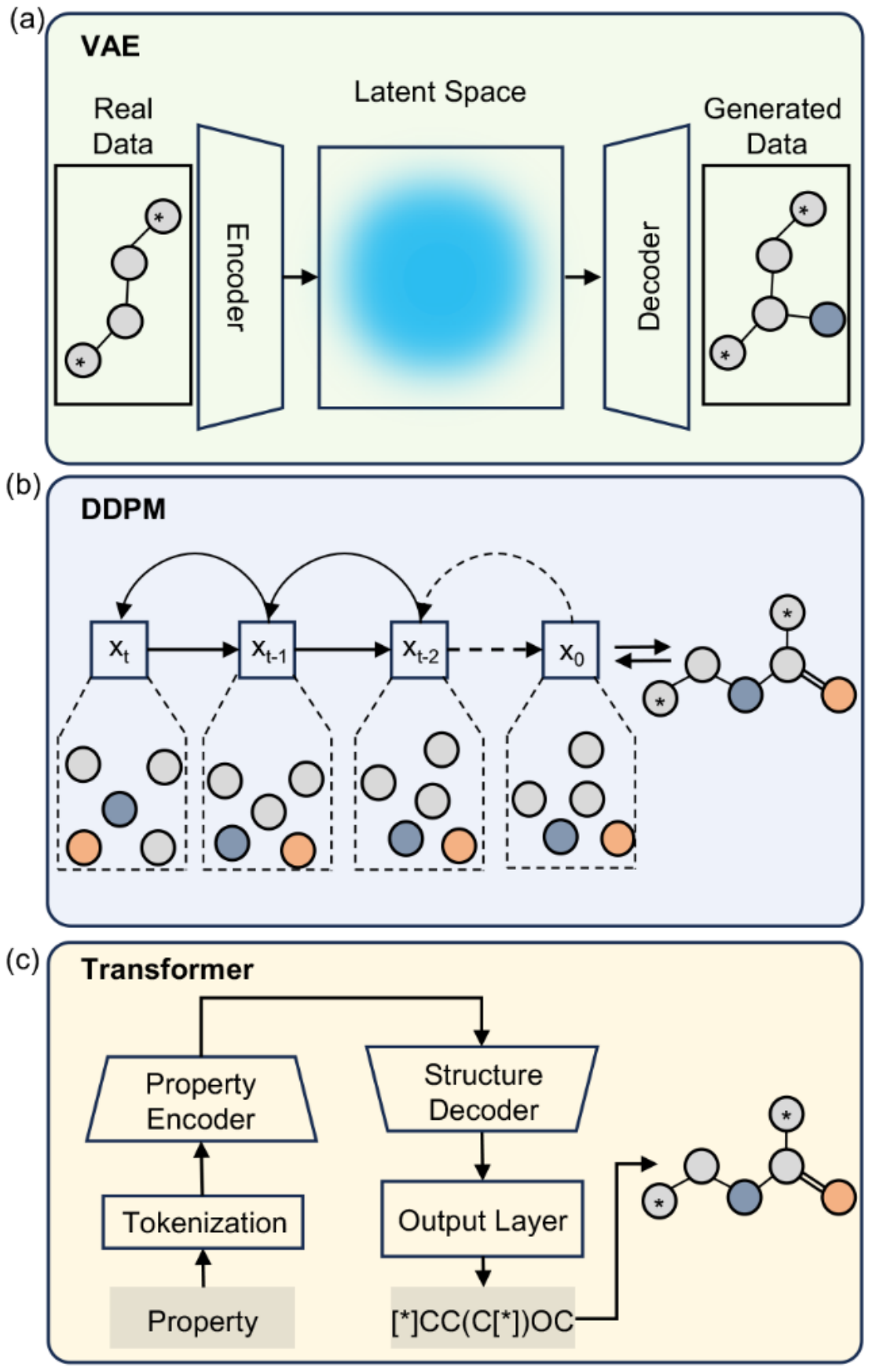

**Figure 4**. **Representative generative models for polymer design**. (a) Variational Autoencoder (VAE). Encodes polymers into a continuous latent space, where new structures are generated by sampling latent vectors and decoding them into valid representations. (b) Denoising Diffusion Probabilistic Model (DDPM). Consists of a forward diffusion process (curved arrow) adding noise and a reverse denoising process (straight arrow) that gradually reconstructs polymer structures from Gaussian noise. (c) Transformer-based model. Learns polymer sequence representations and generates polymers through autoregressive decoding.

The VAE was the first generative framework applied to polymer design, marking a key milestone in establishing the foundation for data-driven inverse design. Its main concept is to map complex molecular structures into a continuous latent representation, enabling both controlled reconstruction and guided generation (**Figure 5 (a)**). By sampling within the latent space, VAE can generate new polymers with similar chemical features.[61] In polymer generation, VAE models commonly employ graph-based representations in which polymers are encoded as node and adjacency matrices (**Figure 5(b)**).[62]

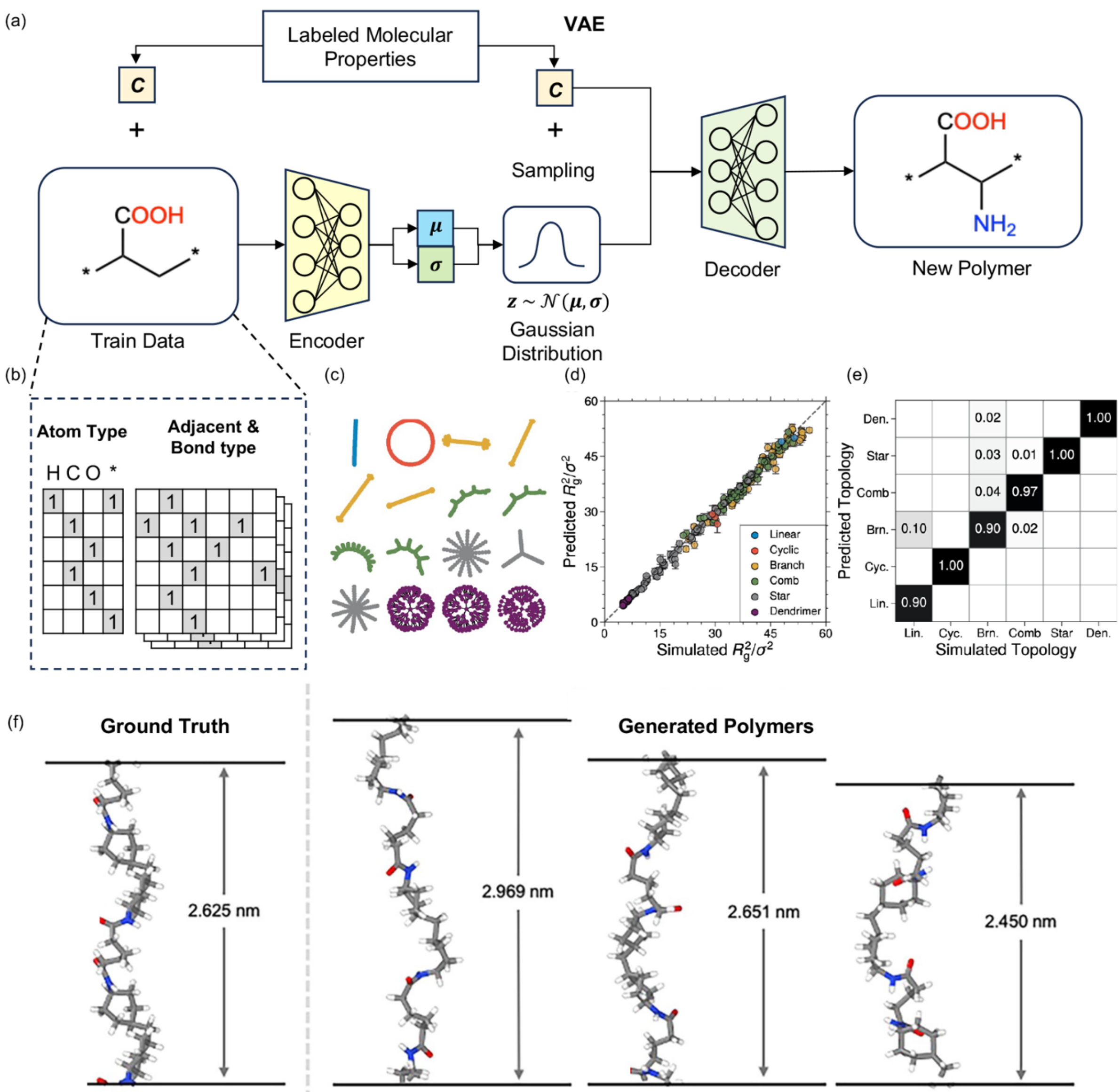

**Figure 5**. **Applications of variational autoencoders (VAEs) in polymer generation**. (a) Schematic illustration of the working principle of a VAE. (b) Graph-based encoding of polymer repeating units, represented using node and adjacency matrices derived from SMILES strings. (c) Generation of polymers with diverse topological architectures using the TopoGNN framework. (d–e) Comparison between predicted and simulated structural or property values for TopoGNN-generated polymers. (f) Multiple 3D conformations generated by polyGen for the same polymer repeating unit. Panel (c–e) reproduced from ref. [63] under the CC BY-NC-ND 4.0 license; Panel (f) reproduced from ref. [64] under the CC BY 4.0 license.

This approach preserves molecular topology and facilitates the generation of chemically valid and structurally consistent candidates during decoding. In 2020, Batra et al.[65] first applied a VAE to polymer design by introducing polymer-specific syntax constraints and a syntactic tree to encode polymer repeat-unit SMILES. Combined with Gaussian process regression, the model achieved property-conditioned generation for $T_g$ and band gap targets. Although the structural validity of the generated polymers was below 30%, highlighting the challenge of capturing chemical constraints in complex macromolecules, this study provided the first demonstration of the feasibility of inverse design for polymers using latent-space generative models. Subsequent research incorporated graph-based encodings to enhance structural accuracy and chemical validity. In 2021, Gurnani et al.[66] developed PolyG2G, a graph-to-graph VAE that treats polymers as 2D molecular graphs of nodes and edges. The model achieved over 90% validity and generated candidate polymers with improved dielectric constants and breakdown strengths compared to the training data. This demonstrated that graph-based representations can improve chemical realism and reduce invalid outputs.

To further enhance generative capability, researchers explored conditional and topology-aware generative models. In 2024, Jiang et al.[63] introduced TopoGNN, which combines GNN-based encoding with topological descriptors such as branching degree and cyclicity to form a hierarchical latent space. This model is capable of generating a diverse range of polymer topologies, including star-shaped polymers and dendrimers (**Figure 5(c)**). It learned structure reconstruction, topology classification and property prediction simultaneously, enabling generation of polymers with target radii of gyration (**Figure 5(d)**) and specified topological architectures (**Figure 5(e)**), highlighting the potential of VAE for joint topology–property design. In 2025, Jain et al.[67] proposed polyGen, the first framework to generate 3D polymer structures directly. Using repeat-unit graphs as inputs, polyGen produces chemically connected and conformationally diverse 3D single-chain structures (**Figure 5(f)**).

Today, VAE-based polymer generators typically achieve chemical validity exceeding 90%. They can integrate property predictors and target constraints, and even generate 3D polymer structures. VAEs have evolved from random generators to goal-directed design frameworks. Their enduring significance lies not only in providing a foundational generative approach but also in demonstrating that latent-space modeling can invert the structure-property relationship, enabling the predictive design of functional polymers.

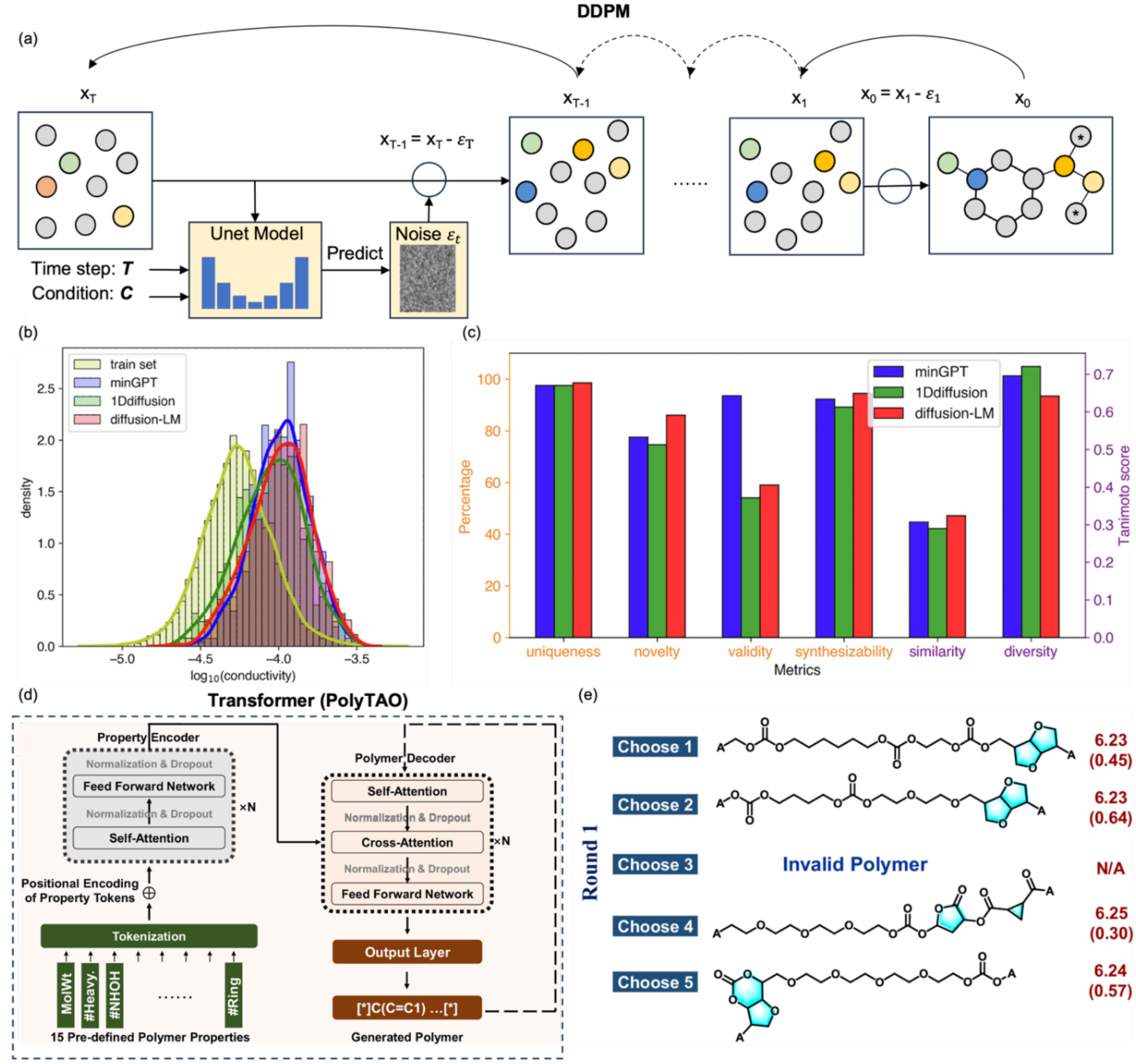

**Figure 6**. **Polymer generation using Denoising Diffusion Probabilistic Models (DDPM) and Transformer architectures**. (a) Workflow of DDPM for generating polymer repeating units. (b) Comparison of $Li^+$ conductivity distributions in the training and generated datasets produced by GPT and DDPM models. (c) Performance of three different generative models. (d) Schematic of the PolyTAO framework, a Transformer-based model that generates polymer repeat-unit SMILES strings conditioned on target properties. (e) Examples of high band gap polymers generated by PolyTAO. Panel (b–c) reproduced from ref.[68] under the CC BY-NC-ND 4.0 license; Panel (d–e) reproduced from ref.[69] under the CC BY-NC-ND 4.0 license.

Following VAE, polymer generation research has advanced from continuous latent representations to probabilistic control and semantic generation. Two mainstream model classes, DDPMs and Transformer-based language models, are now defining this frontier. DDPM generate by gradually adding noise to data and reconstruct them through the reverse process, allowing accurate reconstruction of complex chemical distributions (**Figure 6 (a)**).[70] Yang et al.[68] first applied a diffusion language model to polymer design, exploring its potential for polymer electrolytes. They compared two diffusion paradigms. One approach (1Ddiffusion) adds and removes noise directly in the discrete token sequence. The other approach (diffusion-LM) encodes discrete tokens into a continuous embedding space, performs diffusion and denoising within that space. When conditioned on ionic conductivity labels, the model generated polymers with significantly higher average $Li^+$ conductivity than the training set (**Figure 6 (b)**), confirming the capability of diffusion models for property-targeted generation. However, 1Ddiffusion often disrupted SMILES syntax, yielding around 80% valid structures, while the diffusion-LM achieved higher validity but required multiple denoising iterations per sample, resulting in high computational cost (**Figure 6 (c)**). DDPMs therefore remain in an exploratory phase, with future improvements expected in balancing chemical validity and efficiency.

In contrast, Transformer-based models have demonstrated higher compatibility with polymer sequence data. The sequential nature of polymer SMILES allows Transformers to capture long-range dependencies among fragments through attention mechanisms, enabling semantic understanding of polymerization patterns. Qiu et al. [69] developed PolyTAO, a milestone model that uses a Transformer encoder–decoder to generate polymers conditioned on physical and chemical properties (**Figure 6 (d)**). Trained on nearly one million polymer structure-property pairs, PolyTAO achieved 99.27% chemical validity and an average $R^2$ of 0.96 across 15 predicted properties. It can generate polymers with desired properties such as high band gap, as illustrated in **Figure 6 (e)**. These results highlight the promise of Transformer-based models for inverse polymer design. Currently, verifying the accuracy and reliability of generative model results through experimental data has become increasingly important. Sahu et al.[71] developed POLYT5, a T5-based encoder–decoder model pretrained on more than 100 million polymer structures. Then, the model was applied to dielectric polymer design, followed by large-scale screening and experimental validation of selected materials. The synthesized polymers exhibited thermal properties and band gaps consistent with the predicted values. This study shows that generative models should be evaluated not only by the validity and novelty of their outputs, but also by synthetic accessibility and the agreement between predicted and experimentally measured properties.

In summary, VAE remains the most mature class of generative models in polymer research. Transformer approaches have progressed from early studies to polymer generation and synthesis-oriented workflows, whereas the application of DDPMs to polymers remains at an exploratory stage.

VAE is particularly well-suited for property optimization in low-dimensional latent spaces, for example, via Bayesian optimization, or when interpretability of the generative process is important. In contrast, Transformer-based models are particularly suitable for large sequential datasets where additional information needs to be incorporated, especially in tasks that do not require 3D structure generation. This is due to their flexibility in handling complex input formats and natural compatibility with one-dimensional sequence generation. Finally, when 3D structural generation is essential, for instance, in conformer-sensitive tasks, DDPMs offer a promising solution. Although their application to polymers is still limited, DDPMs have already demonstrated strong performance in molecular conformation generation for small-molecule drug discovery.[72, 73]

## ONLINE PLATFORMS FOR POLYMER

With the rapid progress of AI in polymer science, the focus of research has expanded from algorithm development to practical implementation. Although numerous predictive or generative models have been proposed, their deployment and data preprocessing remain time-consuming, limiting their widespread use in polymer research. To address this bottleneck, several online platforms have been developed in recent years. These platforms integrate data, models, and prediction functions within a user-friendly interface, enabling high-throughput workflows from structure generation to property prediction with minimal effort.

Representative platforms include CEMP[29], Polymer Genome[38], POLYMAT[74], AI plus Polymers[75], and PolyID[76] (**Table 2**). CEMP focuses on homopolymer property prediction and polymer structure generation, integrating the OpenPoly database; Polymer Genome provides property prediction for both homopolymers and copolymers; POLYMAT supports radical polymerization systems by predicting reactivity ratios and rate constants; AI plus Polymers targets the mechanical, thermal, and electrical performance of thermosetting and thermoplastic polymers; and PolyID allows users to perform virtual polymerization and property prediction from monomer inputs, offering high adaptability across diverse polymerization scenarios. Despite differences in functionality, all platforms share user-friendly interfaces, cloud-based computation and easy accessibility, which significantly lower the barrier to using AI for polymer analysis and material screening.

Beyond virtual prediction and screening, integrating AI models with automated experimentation provides a practical route for translating computational designs into experimentally validated materials.

Wu et al.[77] combined LLM-assisted literature mining with a robotic laboratory. The system designed and synthesized electrochromic polymers with previously unreported target color values within 72 h. This framework established a closed-loop workflow that linked knowledge extraction and inverse design with robotic synthesis and optical characterization. Future intelligent platforms for polymer require not only online high-throughput screening and synthesis guidance but also automated laboratories capable of conducting batch experiments.

**Table 2. Representative online platforms for polymer property prediction.**

| Platform | Link | Predictable Properties | Description |
| --- | --- | --- | --- |
| CEMP[29] | https://cleanenergymaterials.cn/polymer | $T_g$, $T_m$, Tensile strength, Young's modulus, Dielectric constant, etc. | Homopolymer property prediction and polymer generation; integrated with the OpenPoly database. |
| Polymer Genome[38] | http://www.polymergenome.org | Band gap, Dielectric constant, Refractive index, Atomization energy, $T_g$, Density, etc. | Property prediction for both homopolymers and copolymers. |
| POLYMAT[74] | https://polymatai.pythonanywhere.com | Reactivity ratios, $T_g$, Solubility, etc. | Focuses on radical polymerization kinetics. |
| AI plus Polymers[75] | http://47.100.230.202/ | Thermal, mechanical, and electrical properties of thermosets and thermoplastics | Serves as a database and prediction platform for thermoset and thermoplastic properties. |
| PolyID[76] | https://polyid.nrel.gov | $T_g$, Tm, Density, Young's modulus, Gas permeability | Enables virtual polymerization and property prediction based on user-defined monomer inputs. |

## CONCLUSION

This review outlines a roadmap toward the realization of a polymer genome by integrating four essential aspects: open and standardized databases, unified encoding strategies, advanced generative models and accessible online platforms. Collectively, these developments are shifting polymer research from empirical exploration to intelligent and data-driven discovery.

In terms of databases, the traditional paradigm of manual data extraction is being replaced by automated and efficient acquisition through LLMs. However, the reliability of LLM-derived data hinges critically on the implementation of rigorous data-cleaning and validation protocols to ensure high confidence and reproducibility. Regarding polymer encoding, representation learning is shifting from single-modality to multimodal strategies. Researchers are also exploring unified encoding strategies that can represent both homopolymers and copolymers. A unified encoding strategy improves the consistency of downstream modeling and enables more effective comparison and transfer between different polymer systems. In property prediction, the field is transitioning from traditional machine learning based on hand-crafted descriptors/fingerprints to deep learning architectures, which can automatically extract features from structural data. Furthermore, the emergence of generative models introduces a new strategy for inverse polymer design, enabling the targeted generation of polymers with desired properties by learning structure-property relationships in reverse. Finally, the development of open and integrated online platforms allows researchers to reuse and extend the existing resources efficiently.

Overall, these four aspects collectively define a scalable framework for AI-driven polymer discovery, thus ensuring the continual innovation in polymer science.

## ACKNOWLEDGMENTS

We appreciate the support of the National Natural Science Foundation of China (92372126, 52373203), the Excellent Young Scientists Fund Program, the AI for Science Foundation of Fudan University (FudanX24AI014) and the Fundamental Research Funds for the Central Universities (20720250005).

## TOC

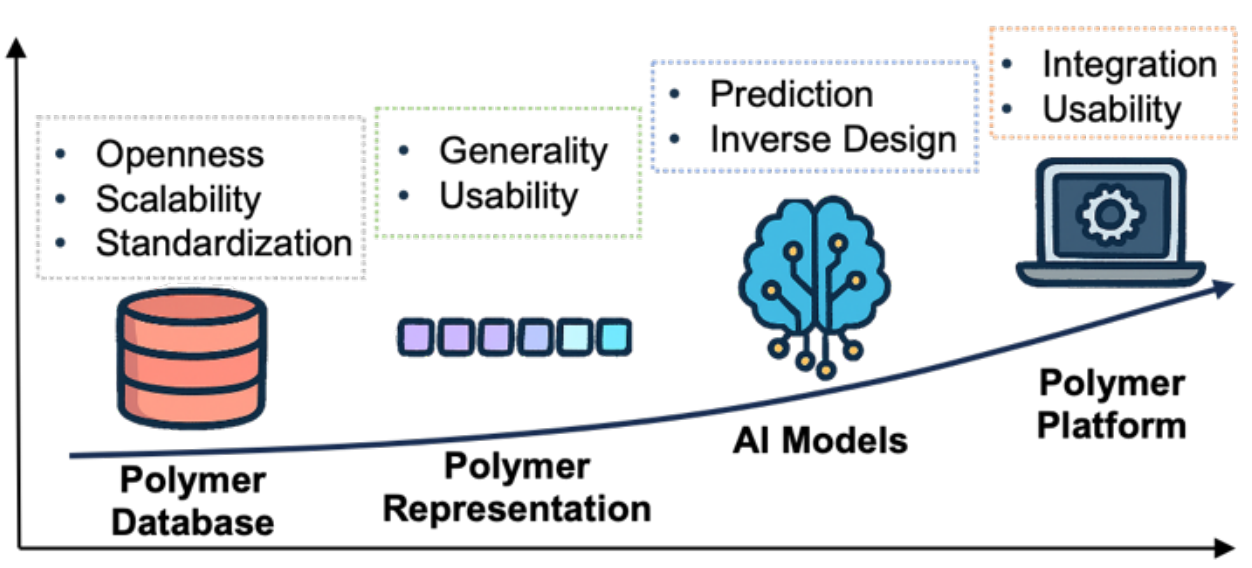

Openness
Scalability
Standardization
Generality
Usability
Prediction
Inverse Design
Integration
Usability
Polymer Database
Polymer Representation
AI Models
Polymer Platform